\documentstyle[twoside,fleqn,epsfig]{article}

\def\Journal#1#2#3#4{{#1} {#2} (#3) #4 }

\def\APP{\em Astrop. Phys.}

\def\PLB{{\em Phys. Lett.} B}

\def\PRD{{\em Phys. Rev.} D}

\def\be{\begin{equation}}
\def\ee{\end{equation}}
\def\bea{\begin{eqnarray}}
\def\eea{\end{eqnarray}}
\newcommand{\lsim}{\mathrel{\mathop{\kern 0pt \rlap
  {\raise.2ex\hbox{$<$}}}
  \lower.9ex\hbox{\kern-.190em $\sim$}}}
\newcommand{\gsim}{\mathrel{\mathop{\kern 0pt \rlap
  {\raise.2ex\hbox{$>$}}}
  \lower.9ex\hbox{\kern-.190em $\sim$}}}

\newcommand{\AmS}{{\protect\the\textfont2
  A\kern-.1667em\lower.5ex\hbox{M}\kern-.125emS}}
\normalsize
\begin{document}

\begin{flushright}
{\bf ROM2F/2007/05\\}
{\bf to appear in Int. J. Mod. Phys. A\\}
\end{flushright}

\vspace{-0.5cm}

\normalsize

\baselineskip=0.65cm
\vspace*{0.5cm}

\begin{center}
\Large \bf
On electromagnetic contributions in WIMP quests\\
\rm
\end{center}

\vspace{-0.2cm}

\vspace{0.5cm}
\normalsize

\noindent \rm R.\,Bernabei,~P.\,Belli,~F.\,Montecchia,~F.\,Nozzoli

\noindent {\it Dip. di Fisica, Universit\`a di Roma ``Tor Vergata"
and INFN, sez. Roma ``Tor Vergata", I-00133 Rome, Italy}  

\vspace{3mm}

\noindent \rm F.\,Cappella, A.\,Incicchitti,~D.\,Prosperi

\noindent {\it Dip. di Fisica, Universit\`a di Roma ``La Sapienza"
and INFN, sez. Roma, I-00185 Rome, Italy}

\vspace{3mm}

\noindent \rm R.\,Cerulli

\noindent {\it Laboratori Nazionali del Gran Sasso, INFN, Assergi, Italy}

\vspace{3mm}

\noindent \rm C.J.\,Dai,~H.L.\,He,~H.H.\,Kuang,~J.M.\,Ma,~X.D.\,Sheng,~Z.P.\,Ye\footnote{also:
University of Jing Gangshan, Jiangxi, China}

\noindent {\it IHEP, Chinese Academy, P.O. Box 918/3, Beijing 100039, China}

\vspace{0.5cm}
\normalsize

\begin{abstract}

The effect pointed out by A. B. Migdal in the 40's (hereafter named Migdal effect)
has so far been usually neglected in the direct searches for WIMP Dark Matter candidates. 
This effect consists in the ionization and the excitation of bound atomic 
electrons induced by the recoiling atomic nucleus.
In the present paper the related theoretical arguments are developed and  
some consequences of the proper accounting for this effect are 
discussed by some examples of practical interest.
 
\end{abstract}



\section{Introduction}

The Migdal effect is known since long time and is described both in devoted papers \cite{migdalp} 
and in textbooks \cite{migdaltb}; it has also been recently addressed for the Dark Matter 
(DM) field in ref. \cite{vergados}. This effect consists in the ionization and the excitation of bound atomic
electrons induced by the presence of a recoiling atomic nucleus.
In this paper, it will be accounted in the case of the WIMP-nucleus
elastic scattering. In fact, since the recoiling nucleus can  "shake off" some of the atomic electrons,  
an electromagnetic contribution is present together with a recoil signal in the analysis of
DM direct searches (with whatever approach) when interpreted in terms of WIMP candidates.
Since this contribution is not quenched, one can expect that this part (usually unaccounted)
can play a role as well.

In this paper the related theoretical framework is developed and some of the previous 
corollary analyses for WIMP candidates \cite{RNC,ijmd,epj06} from the DAMA/NaI 
annual modulation results (total exposure of 107731 kg $\times$ day)
are used for template purpose. We just remind that, 
in order to investigate in a model independent way the presence of DM particle component(s) 
in the galactic halo, DAMA/NaI \cite{Nim98,Sist,RNC,ijmd} has exploited over seven annual cycles the DM 
annual modulation signature, achieving a 6.3 $\sigma$ C.L. model independent
evidence \cite{RNC,ijmd}.
Some of the many possible corollary quests for the candidate 
particle(s) have also been carried out so far mainly focusing
various possibilities for the class of DM candidate particles named WIMPs and for the class of light bosons 
\cite{RNC,ijmd,ijma,epj06}; other corollary quests are also available in literature, such as e.g.
in refs. \cite{Bo03,Bo04,Botdm,khlopov,Wei01,foot,Saib}, and many other scenarios can be considered as well.

\section{Atomic effects due to nuclear recoils}

The possible excitation and ionization of the atom by a recoiling nucleus --
induced by a WIMP-nucleus elastic scattering -- give rise to a certain quantity of
electromagnetic radiation made of the escaping electron and of X-rays and/or Auger electrons
arising from the rearrangement of the atomic shells. This radiation is fully contained in a detector of suitable size.

Since the WIMP-nucleus interaction is expected to have a very short range
(e.g. being mediated by very heavy particles with mass $M$),
the duration of the collision 
($\sim \hbar/M \lsim 10^{-26}$ s) is negligible with respect 
to the electron orbit periods and to $R_a/V_A$,
where $R_a$ is the atomic size and $V_A$ is the nucleus velocity after the interaction.
Thus, the interaction can be considered as instantaneous and 
a semiclassical description of the recoil process can be applied following the Migdal 
approach \cite{migdalp,migdaltb}: 
the target nucleus is assumed to be at rest for $t<t^*$ and at 
$t=t^*$ it suddenly acquires the $\vec{V}_A$ velocity
because of the interaction.
Therefore, after the collision the wave function of the
electronic states at $t=t^*$ can be approximated as \cite{migdaltb}:
\begin{equation}
\Psi'_i = \left[ e^{-i\vec{Q}_A \cdot \sum_\alpha \vec{r}_\alpha} \right] \cdot
\Psi_i(\vec{r}_1,\vec{r}_2,...) .
\label{eq:M1}
\end{equation}
In this relation:  i) $\Psi_i(\vec{r}_1,\vec{r}_2,...)$ is their wave function in the 
rest frame of the nucleus before the interaction;
ii) $\vec{r}_\alpha$ is the coordinate vector of the $\alpha-th$ electron;
iii) $\vec{Q}_A = m_e \vec{V}_A$ with $m_e$ electron mass.

From eq. (\ref{eq:M1}) the probability $P_{fi}$ to reach the intrinsic state 
$|\Psi_f>$ starting from an initial state $|\Psi'_i>$ can be written as:
\begin{equation}
P_{fi} =\left| <\Psi_f|\Psi'_i> \right|^2 = \left| <\Psi_f|e^{-i\vec{Q}_A \cdot \sum_\alpha \vec{r}_\alpha}|\Psi_i> \right|^2  .
\label{eq:M2}
\end{equation}

In the following, the case of the transition of an electron from one level
to another and the case of its transition to the continuum (that is, the ionization of the atom) 
are separately analysed applying
the reasonable approximation that multiple transition/ionization processes can safely be neglected.

\subsection{Calculation of the excitation probability and profile}

The total transition probability can be calculated from the single electron 
transition probability, $P^1_{fi}$, by exploiting the   
mean field approximation. In particular, we consider that the wave function of a single
electron does not depend on the coordinates of the other ones.
Thus, the probability $P^1_{fi}$ of the transition of a 
single electron from an initial state $|\psi_{i}>$ to a final bound state 
$|\psi_{f}>$ is simply: 

\begin{equation}
P^1_{fi} = \left| <\psi_{f}|e^{-i\vec{Q}_A \vec{r}}|\psi_{i}> \right|^2 ,
\label{eq:M5}
\end{equation}
where $\vec{r}$ is the coordinate vector of the considered electron.

The term $e^{-i\vec{Q}_A  \vec{r}}$ 
can safely be expanded in series by applying the dipole approximation.
In fact, since $V_A$ is at maximum 
of the order of the impinging WIMP velocity (less than the Galaxy escape velocity: $\sim 2 \cdot 10^{-3}c$)
and $R_a \sim 10^{-10}$ m,
we can write $\epsilon = Q_AR_a \simeq \frac{m_e V_A R_a}{197 MeV fm} \lsim 1$.
Thus, considering that in practice: $\psi_{i}(\vec{r}) \sim 0$ for $r > R_a$
and $ \vec{Q}_A  \vec{r}  \le Q_Ar$, the amplitude in eq. (\ref{eq:M5}) can be re-written as:
\begin{eqnarray}
&<\psi_{f}|e^{-i\vec{Q}_A  \vec{r}}|\psi_{i}>
\simeq 
<\psi_{f}| (1 - i \vec{Q}_A  \vec{r} -
 \frac{1}{2}( \vec{Q}_A  \vec{r})^2 + ... )
|\psi_{i}> 
\label{eq:M6}
\end{eqnarray}
retaining only the leading orders.

\begin{figure} [!ht]
\centering
\includegraphics[width=185pt] {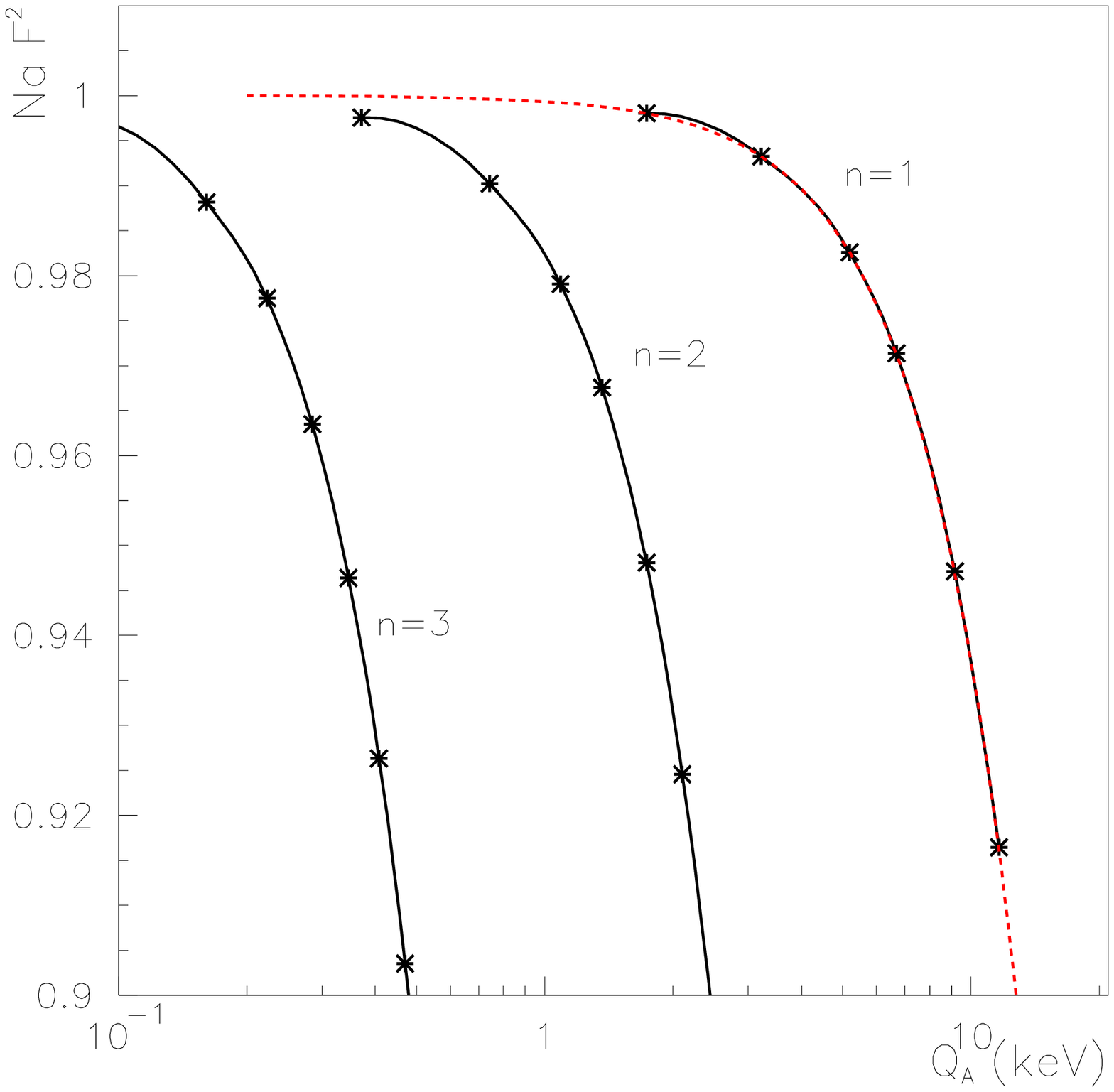}
\includegraphics[width=185pt] {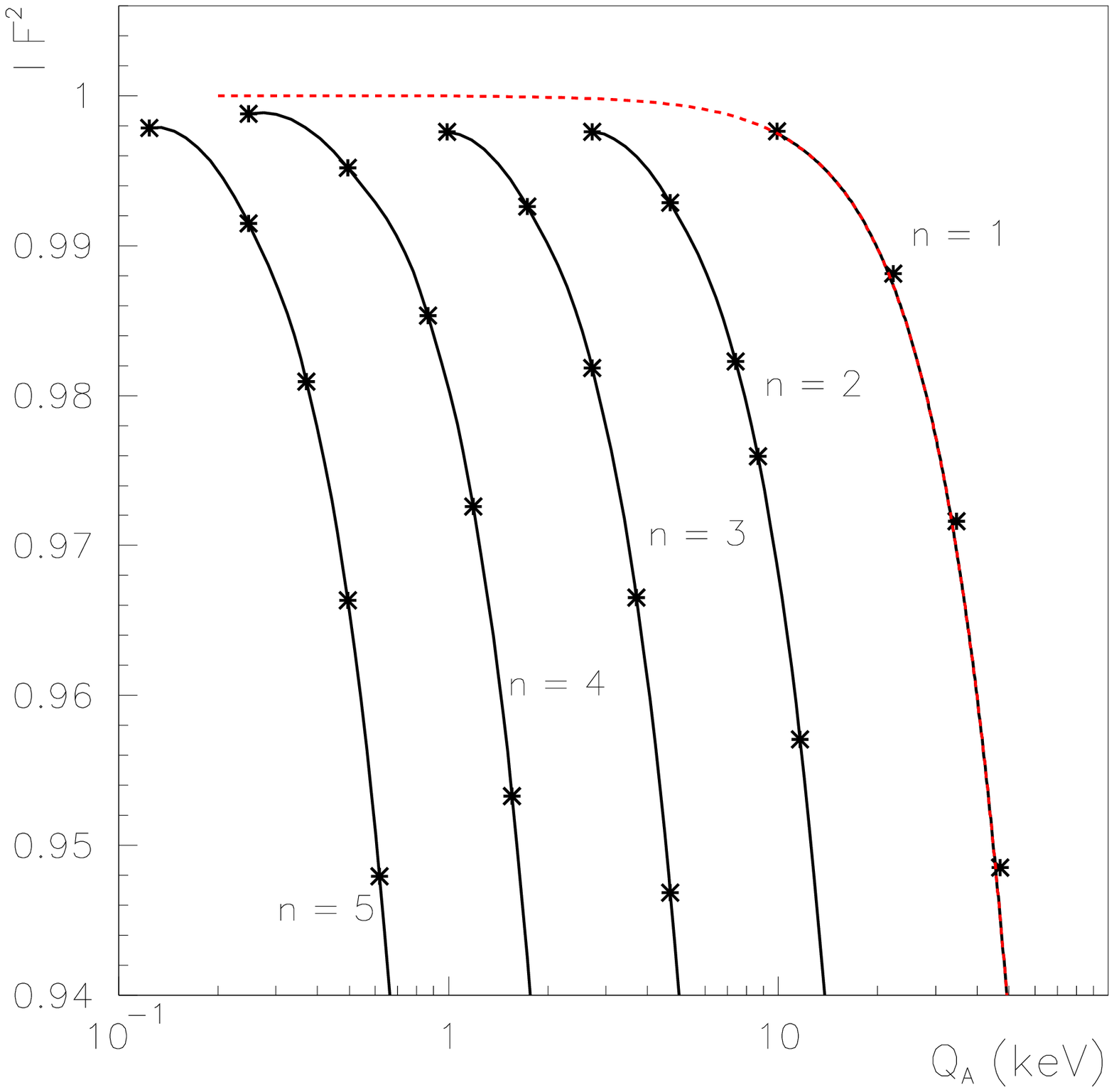}
\vspace{-0.6cm}
\caption{Atomic Form Factors from \cite{rtab} averaged over the shell for 
various principal quantum numbers of Sodium (left) and Iodine (right) atoms.
For the shell 1s the Form Factors calculated by means of the hydrogenic wave functions 
(dashed lines) are also reported for comparison.}
\label{fg:FF}
\vspace{-0.3cm}
\end{figure}

As first, the probability for a single electron to retain the same state can be written as:
\begin{eqnarray}
P^1_{ii} & = & \left| <\psi_{i}|e^{-i\vec{Q}_A \vec{r}}|\psi_{i}> \right|^2 = F^2_i\nonumber \\
 & \simeq & \left| <\psi_{i}|(1 - i \vec{Q}_A  \vec{r} - ...)|\psi_{i}> \right|^2 \sim
1 - O(\epsilon^2) \lsim 1 ,
\label{eq:M7}
\end{eqnarray}
where $F_{i}$ is the atomic Form Factor of the orbital described by the $\psi_{i}$ 
wave function. It can be derived
from the Rayleigh scattering database (RTAB) \cite{rtab}, 
where the atomic Form Factors for the levels of the various atoms are reported.
Some examples of the used atomic Form Factors for the Sodium
and Iodine atoms are reported in Fig. \ref{fg:FF};
for the shell 1s the atomic Form Factors calculated by means of the hydrogenic wave functions 
are also reported for comparison, showing a good agreement.

On the other hand, for the transitions ($f \neq i$) one gets:
\begin{equation}
P^1_{fi} \simeq \left| <\psi_{f}|(1 - i \vec{Q}_A  \vec{r} - ...)|\psi_{i}> \right|^2
\simeq \left| <\psi_{f}| \vec{Q}_A  \vec{r} |\psi_{i}> \right|^2 \sim O(\epsilon^2) \ll 1
\label{eq:M8}
\end{equation}
because of the states orthogonality. 

If we define $\theta$ as the angle 
between the vectors $\vec{Q}_A$ and $\vec{r}_{fi} = <\psi_{f}| \vec{r} |\psi_{i}>$, 
this transition probability can be re-written as:
\begin{equation}
P^1_{fi} \simeq  cos^2(\theta) Q_A^2 \left|r_{fi}\right|^2 \; . 
\label{eq:M14}
\end{equation}
Since the atoms are unpolarized, $cos^2(\theta) $ can be replaced by its average: 1/3, that is

\begin{equation}
P^1_{fi} \simeq  \frac{1}{3} Q_A^2 \left|r_{fi}\right|^2 \; .
\label{eq:M14_2}
\end{equation}

The $r_{fi}$ matrix elements are connected with the standard dimensionless 
oscillator strengths, $f_{fi}$, by means of (as usual, here: $\hbar=1$):
\begin{equation}
f_{fi}=\frac{2m_e \omega_{fi}}{3}|r_{fi}|^2 
\label{eq:M14os}
\end{equation}
where $\omega_{fi}$ is the energy of the transition and,
in conclusion, the transition probability can be expressed as:
\begin{equation}
P^1_{fi} \simeq  \frac{Q_A^2 f_{fi}} {2 m_e \omega_{fi}}\; .
\label{eq:M14_a}
\end{equation}

The oscillator strengths for several atoms can also be derived from  the Rayleigh scattering database \cite{rtab}.
Some examples of averaged oscillator strengths, calculated on this basis for various atoms,
are reported in Table \ref{tb:stren} together with some other evaluations available in literature.

\begin{table}[!ht]
\vspace{-0.4cm}
\caption{Some examples of the averaged oscillator strengths -- calculated 
from the Rayleigh scattering database \cite{rtab} -- for various atoms
and transitions, together with some other evaluations available in literature.}
\begin{center}
\begin{tabular}{|c|c|c|c|}
\hline \hline
 Atom &  Transition         & $f_{fi}$ from \cite{rtab}  & $f_{fi}$ from literature \\
\hline 
  H   & $1s \rightarrow 2p$ & 0.416                      &  0.416 \cite{ppa} \\
  H   & $1s \rightarrow 3p$ & 0.079                      &  0.079 \cite{ppa} \\
  H   & $1s \rightarrow 4p$ & 0.029                      &  0.029 \cite{ppa} \\
 Na   & $3s \rightarrow 2p$ & -0.042                     & -0.043 \cite{HTI} \\
 Na   & $3s \rightarrow 3p$ & 0.968                      &  0.983 \cite{HTI} \\
 Li   & $2s \rightarrow 2p$ & 0.762                      &  0.753 \cite{HTI} \\
  K   & $4s \rightarrow 4p$ & 1.045                      &  1.02  \cite{HTI} \\
\hline \hline
\end{tabular}
\end{center}
\label{tb:stren}
\vspace{-0.4cm}
\end{table}

\vspace{0.3cm}

Let us now consider the general case. 
The probability of a single electron in the state $i$ to have a transition
to a free bound level (excitation) can be derived by summing over all the available 
possible states:
\begin{equation}
P^1_{bound,i} \simeq \sum_{j \in free \; bound \; states}
|<\psi_{j}| \vec{Q}_A
\vec{r}|\psi_{i}>|^2 ,
\label{eq:M8e6}
\end{equation}
neglecting the probability that other electrons
fill the $j$-th state in the meantime. 
After an excitation, the atom returns to the ground state emitting
X-rays and/or Auger electrons. The total energy released by the process 
is equal to the transition energy and it depends on the energy level of the final state.
Considering that all the free bound levels are within at maximum few eV,
we can safely neglect the spread of the excitation energies.
Thus, the excitation profile of the $i$-th electron can simply be
assumed to be a Dirac delta function centered
around the average excitation energy, $\langle E \rangle_i$;
this is generally less than few eV below the ionization threshold.
Therefore, considering the excitations of the single $i$-th electron, 
the differential distribution of the electromagnetic part of the detected 
energy\footnote{It is the sum of all the energies 
of the X-rays and Auger electrons, while in case of the ionization 
process -- see later -- also the energy of the electron 
escaping from the atom contributes.}, $E_{em}$,
for a given energy, $E_{0}$, 
provided by the WIMP to the nucleus at time $t^*$,
is given by:
\begin{equation}
\frac{dN_{excitation,i}}{dE_{em}} (E_{em}|E_{0})=
P^1_{bound,i} \delta(E_{em} - \langle E \rangle_i)
\; .
\label{eq:s1_exc}
\end{equation}

\subsection{Calculation of the ionization probability and profile}

The probability of a single electron in the state $i$ to have a transition
to the continuum (ionizing the atom), $P^1_{ion,i}$,
can be derived as the difference between the unity and
the probability of all the other possible processes. They are the following:
i)  the electron has no transition ($P^1_{ii} = F_{i}^2$);
ii) the electron inter-exchanges the states with another electron ($P^1_{exc,i}$);
iii) the electron excites to a free bound level ($P^1_{bound,i}$). Hence:
\begin{equation}
P^1_{ion,i} = 1 - P^1_{ii} - P^1_{exc,i} - P^1_{bound,i}
\label{eq:M8e5}
\end{equation}

Taking into account the Pauli exclusion principle, the transition
to an occupied state $j$ is possible only if the $j$-th electron
goes to another state. Thus, the exchange probability can be estimated
by the expression:
\begin{equation}
P^1_{exc,i} \simeq \sum_{j \in occupied \; bound \; states}
|<\psi_{j}| \vec{Q}_A
\vec{r}|\psi_{i}>|^2 (1-F_{j}^2) 
\label{eq:M8e7}
\end{equation}
having neglected the probability that other electrons
fill the $j$-th state in the meantime. 

Unlike the excitation, the differential distribution, $\frac{dN_{ionization,i}}{dE_{em}}$, 
considering the ionization due to the single $i$-th electron gives a continuum spectrum:
\begin{equation}
\frac{dN_{ionization,i}}{dE_{em}} (E_{em}|E_{0})=
P^1_{ion,i} D^{i}_{ion}(E_{em}) \; ,
\label{eq:s1_ion}
\end{equation}
where the ionization profile $D^{i}_{ion}(E_{em})$ is evaluated in the following (see later).

\begin{table}[t]
\caption{Estimated probabilities of excitation and ionization by Migdal effect during
a 10 keV kinetic energy Na recoil or during a 33.3 keV kinetic energy I recoil.
Although the calculated probabilities 
are quite small, the unquenched nature of the electromagnetic
contribution, the behaviour of the energy distribution for nuclear recoils 
induced by WIMP-nucleus elastic scatterings, etc. can
give an appreciable impact at low WIMP masses; see later.}
\begin{center}
\begin{tabular}{|c|c|c|c|c|c|c|}
\hline \hline
Atom&Shell & $P^1_{ii}$  & $P^1_{exc,i}$ & $P^1_{bound,i}$ & $P^1_{ion,i}$ & $<E>_i$ (keV)\\
\hline
   &1s & 0.99985 & $1.7 \cdot 10^{-7}$ & $7.8 \cdot 10^{-7}$ & $1.5 \cdot 10^{-4}$ & 1.062 \\
Na &2s & 0.99595 & $1.2 \cdot 10^{-5}$ & $2.7 \cdot 10^{-5}$ & $4.0 \cdot 10^{-3}$ & 0.062 \\
   & 2p & 0.99595 & $4.1 \cdot 10^{-6}$ & $1.5 \cdot 10^{-4}$ & $3.9 \cdot 10^{-3}$ & 0.033 \\
\hline
&1s & 0.99999 & $2.9 \cdot 10^{-10}$ & $2.8 \cdot 10^{-10}$ & $3.3 \cdot 10^{-6}$ & 33.166 \\
&2s & 0.99996 & $8.4 \cdot 10^{-9}$  & $7.3 \cdot 10^{-9}$  & $4.5 \cdot 10^{-5}$ & 5.161\\
&2p & 0.99996 & $9.8 \cdot 10^{-9}$  & $8.9 \cdot 10^{-9}$  & $4.5 \cdot 10^{-5}$ & 4.687\\
&3s & 0.99966 & $1.9 \cdot 10^{-7}$  & $1.1 \cdot 10^{-7}$  & $3.4 \cdot 10^{-4}$ & 1.054\\
&3p & 0.99966 & $1.8 \cdot 10^{-7}$  & $1.5 \cdot 10^{-7}$  & $3.4 \cdot 10^{-4}$ & 0.894\\
I&3d & 0.99966 & $6.6 \cdot 10^{-8}$  & $3.5 \cdot 10^{-8}$  & $3.4 \cdot 10^{-4}$ & 0.635\\
&4s & 0.99730 & $5.5 \cdot 10^{-6}$  & $1.3 \cdot 10^{-6}$  & $2.7 \cdot 10^{-3}$ & 0.190\\
&4p & 0.99730 & $6.8 \cdot 10^{-6}$  & $3.8 \cdot 10^{-6}$  & $2.7 \cdot 10^{-3}$ & 0.138\\
&4d & 0.99730 & $6.5 \cdot 10^{-6}$  & $1.2 \cdot 10^{-5}$  & $2.7 \cdot 10^{-3}$ & 0.058\\
&5s & 0.98102 & $2.5 \cdot 10^{-4}$  & $2.5 \cdot 10^{-5}$  & $1.9 \cdot 10^{-2}$ & 0.019\\
&5p & 0.98102 & $8.5 \cdot 10^{-5}$  & $7.3 \cdot 10^{-3}$  & $1.2 \cdot 10^{-2}$ & 0.009 \\
\hline \hline
\end{tabular}
\end{center}
\label{tb:prob}
\end{table}  

In Table \ref{tb:prob} some numerical estimates of the probabilities of the Migdal effect, 
discussed in the previous and in the present sections, are given.
It is worth to note that, although these probabilities 
are quite small, the electromagnetic unquenched nature of the
contribution, the behaviour of the energy distribution for nuclear recoils 
induced by WIMP-nucleus elastic scatterings, etc.
have some impact at low WIMP masses (see later).

Considering the free electron approximation (that is, neglecting the interaction of 
the escaping electron with the nucleus), the final state of the escaping electron 
can be described as a plane wave \cite{Niel01}: 
$\psi_{\vec{p}_f} \simeq \frac { e^{i\vec{p}_f \vec{r}}}{\sqrt{V}}$,
normalized to a volume $V$.
The final state density is: $\rho_f = \frac {Vd^3p_f}{(2\pi)^3}$.

Thus, the ionization probability, for an escaping electron with momentum
$\vec{p}_f$, can be written as:
\begin{eqnarray}
dW_{\vec{p}_fi} & = & \left| <\psi_{\vec{p}_f}|e^{-i \vec{Q}_A
\vec{r}}|\psi_{i}> \right|^2 \rho_f
\simeq \left| \int \frac { e^{-i\vec{p}_f \vec{r}}} {\sqrt{V}}
e^{-i \vec{Q}_A  \vec{r}} \psi_{i}(\vec{r}) d^3r \right|^2
\frac {Vd^3p_f}{(2\pi)^3} 
\simeq \nonumber \\ 
& \simeq &
 \left| \int \frac { e^{-i(\vec{Q}_A+\vec{p}_f) \vec{r}}} {\sqrt{(2\pi)^3}}
\psi_{i}(\vec{r}) d^3r \right|^2
d^3p_f = 
\left| \phi_{i}(\vec{p}) \right|^2 d^3p_f = 
\rho_{i}(\vec{p}) d^3p_f \; ,
\label{eq:M11}
\end{eqnarray}
\noindent where: i)  $\vec{p} = \vec{p}_f + \vec{Q}_A$; ii)
$\phi_{i}(\vec{p})$ is the wave function of the initial state in the momentum space;
iii) $\rho_{i}(\vec{p})$ is the probability density to find 
a momentum $\vec{p}$ when the electron is bound with wave function $\psi_{i}$.
The ionization profile, $\rho_i(\vec{p})$, is normalized to 1, as it can be easily demonstrated:
$ \int \phi^*_i (\vec{p})\phi_i (\vec{p}) 
d^3p_f = \langle \phi_i \vert \phi_i \rangle = 1$.

Finally, for practical purposes $Q_A \ll p_f$; 
in fact, since the maximum energy provided by the WIMP with mass, $m_W$,
and velocity, $v$, to a nucleus with mass, $m_A$, is 
$E_{0,max}=\frac{1}{2} m_W v^2 \frac{4m_Wm_A}{(m_W+m_A)^2}$, the 
maximum velocity is: $V_{A,max}=\frac{2m_W}{m_W+m_A} v \leq 2 v$.
Therefore, $Q_A \ll 3$ keV and can be safely neglected with respect to 
$p_f$ when the electron energy is larger than 10 eV.

The energy distribution of
the escaping electron can be obtained from eq. (\ref{eq:M11}) by introducing the variable 
$E_f = \frac{p_f^2}{2 m_e}$ (escaping energy of the electron) and by integrating 
over the solid angle.
The angular integration can be easily performed by considering an average ionization
profile for all the $Z_{shell}$ electrons in a full shell, where $\rho_{shell}(p_f) 
= \frac{1}{Z_{shell}} \sum_{shell} \rho_i(\vec{p}_f)$ is isotropic.
Thus, we can define the ionization profile of a given shell as a function of the
escaping energy of the electron according to: 
\begin{equation}
D^{shell}_{ion}(E_f) \cdot dE_f
\simeq
4\pi m_e \cdot \sqrt{2m_eE_f} \cdot \rho_{shell}\left(p_f \right)
\cdot dE_f \; ,
\label{eq:finalrho}
\end{equation}

Moreover, in the real cases, the ionization is energetically allowed 
only for $E_b \leq E_f \leq E_{0}$, with $E_b$ binding energy of the electron. 

\begin{figure} [!ht]
\centering
\vspace{-0.8cm}
\includegraphics[width=200pt] {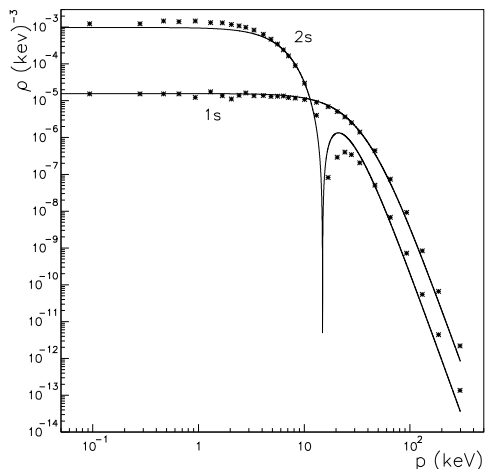}
\vspace{-0.6cm}
\caption{Example of $\rho_{1s}(p)$ and $\rho_{2s}(p)$ for Na atom
as extracted from the Compton profile tables \cite{cprof}
by means of the procedure given in the text (stars).
The solid lines -- shown for comparison --
are the $\rho_{1s}(p)$ and $\rho_{2s}(p)$ analytically calculated 
using hydrogenic wave functions
and accounting for the screening effect of the inner electrons
($Z_{eff} = 10$ and $Z_{eff} = 8$ respectively).}
\label{fg:Nacp}
\end{figure}

In order to obtain the functions $\rho_{shell}(p)$, one can use the 
tables of the Compton profile   
$J_i(p_z)=\int \int \rho_i(\vec{p}) dp_xdp_y$ available in literature \cite{cprof}
for the various shells of atoms \footnote{In fact, 
for isotropic distributions (e.g. for s shell or for full shells) 
the following relation holds \cite{cprof2,cprof3}:
$J_{shell}(p_z)=\int \int \rho_{shell}(p) dp_xdp_y = 2\pi \int_{p_z}^{\infty} p \rho_{shell}(p) dp$
and, hence: $\rho_{shell}(p) = -\frac{1}{2\pi p} \frac{dJ_{shell}(p)}{dp}$.
Thus, the $\rho_{shell}(p)$ values of interest can be evaluated by means of 
the $J_{shell}(p)$ tables given in ref. \cite{cprof}.}.
As examples, $\rho_{1s}(p)$ and $\rho_{2s}(p)$ for Na atom
are reported in Fig. \ref{fg:Nacp}
as extracted from the Compton profile tables \cite{cprof}.
They show a rather good agreement with the values calculated by using the hydrogenic 
wave functions and accounting for the 
screening effect of the inner electrons; these latter ones are also shown in the figure 
for comparison.

\section{Calculation of the expected counting rate}

For each given energy, $E_{0}$, provided by the WIMP to the nucleus at time $t^*$,
the differential distribution, $\frac{dN_i}{dE_{em}}$, considering the transitions
(ionizations/excitations) of the single $i$-th electron, is given by
the sum of the two contributions of eq. (\ref{eq:s1_exc}) and eq. (\ref{eq:s1_ion}).
Moreover, the probability density distribution to
have an electromagnetic release, $E_{em}$, for a given $E_{0}$ value
(considering all the contributions from all
the electrons in the atom) can be written as:%
\begin{equation}
\frac{dN_{tot}}{dE_{em}} (E_{em}|E_{0}) \simeq P^0_{E_{0}} \delta(E_{em}) + \sum_i \frac{P^0_{E_{0}}}{P^0_{i,E_{0}}}
\frac{dN_i}{dE_{em}} (E_{em}|E_{0}) + ... \;  ,
\label{eq:s3}
\end{equation}  
where the term $P^0_{E_{0}}$ is the probability for the whole atom to remain unchanged:
$P^0_{E_{0}} \simeq \prod_i P^0_{i,E_{0}}$,   
and $P^0_{i,E_{0}}= 1-P^1_{ion,i}-P^1_{bound,i}$ is the probability that 
a single electron has no transition either to free bound level or to the continuum.
In eq. (\ref{eq:s3}) the expansion can be stopped at the shown level because the probability
of multiple excitation/ionization is negligible.

Finally, let us rewrite eq. (\ref{eq:s3}) as a function of the detected energy, $E_{det}$, which is 
given by the sum of the recoil detected energy, $E_{fr}$, and of the electromagnetic component
(the relation $\frac{E_{fr}}{q_A} + E_{em} = E_0$ holds):

\begin{equation}
\frac{dN}{dE_{det}} (E_{det}|E_{0}) = \frac{dN_{tot}}{dE_{em}} (E_{em}|E_{0}) \cdot 
\frac{dE_{em}}{dE_{det}} =
\frac{1}{1-q_A} \cdot \frac{dN_{tot}}{dE_{em}} (E_{em}|E_{0}) \; ,
\label{eq:s5}
\end{equation}
where $q_A$ is the nuclear recoil quenching factor for the considered nucleus in the given detector
at the considered energy.

In particular, let us now point out the case of recoils induced by WIMP-nucleus elastic scatterings
under the usual hypothesis that just one component of the dark halo can produce elastic 
scatterings on nuclei.

For every target specie $A$, the expected energy
distribution including the Migdal effect, $\frac{dR^{(M)}_A}{dE_{det}}
(E_{det})$, requires the $E_0$ differential distribution 
produced in the WIMP-nucleus elastic scattering, given in squared brackets: 
\begin{equation}
\frac{dR^{(M)}_A}{dE_{det}}(E_{det}) = \int
\frac{dN}{dE_{det}} (E_{det}|E_{0})
\left[ N_{T}\frac{\rho_{W}}{m_W}\int^{v_{max}}_{v_{min}(E_{0})}
\frac{d\sigma}{dE_{0}}(v,E_{0}) v f(v) dv \right] dE_0 \; . 
\label{eq:s6}
\end{equation}
There: 
i) $N_T$ is the number of target nuclei of A specie; 
ii) $\rho_W = \xi \rho_0$, where
$\rho_0$ is the local halo density and $\xi \leq 1$ 
is the fractional amount of local WIMP density;
iii) $f(v)$ is the WIMP velocity ($v$) distribution in the Earth frame;
iv) $v_{min} = \sqrt{\frac {m_A \cdot E_0}{2 m^2_{WA}}}$ ($m_{WA}$ is 
the reduced mass of the WIMP-nucleus system);
v) $v_{max}$ is the maximal WIMP velocity in the halo
evaluated in the Earth frame;
vi) $ \frac{d\sigma}{dE_{0}}(v,E_{0}) = \left( \frac{d \sigma}{dE_{0}} \right)_{SI}+
\left( \frac{d \sigma}{dE_{0}} \right)_{SD} $,
with $\left( \frac{d \sigma}{dE_{0}} \right)_{SI}$ 
spin independent (SI) contribution and
$\left( \frac{d \sigma}{dE_{0}} \right)_{SD}$ 
spin dependent (SD) contribution.

Finally, the expected differential counting rate as a function of the detected energy, 
$E_{det}$, for a real multiple-nuclei detector (as e.g. the NaI(Tl))
can be easily derived by summing the eq. (\ref{eq:s6}) over the nuclei species
and taking into account the detector energy resolution:

\begin{equation}
\frac{dR^{NaI}}{dE_{det}}(E_{det}) =
\int G(E_{det},E') \sum_{A=Na,I}
\frac{dR^{(M)}_A}{dE'}
(E') dE' \; .
\label{eq:labelmul}
\end{equation}

\noindent The $G(E_{det},E')$ kernel generally has a gaussian behaviour.

Obviously the expected differential counting rate has to be evaluated in given 
astrophysical, nuclear and particle physics scenarios, also requiring
assumptions on all the parameters needed in the calculations and the proper consideration
of the related uncertainties (for some discussions see e.g. \cite{RNC,ijmd,epj06}).

\begin{figure} [!ht]
\centering
\vspace{-0.8cm}
\includegraphics[width=250pt] {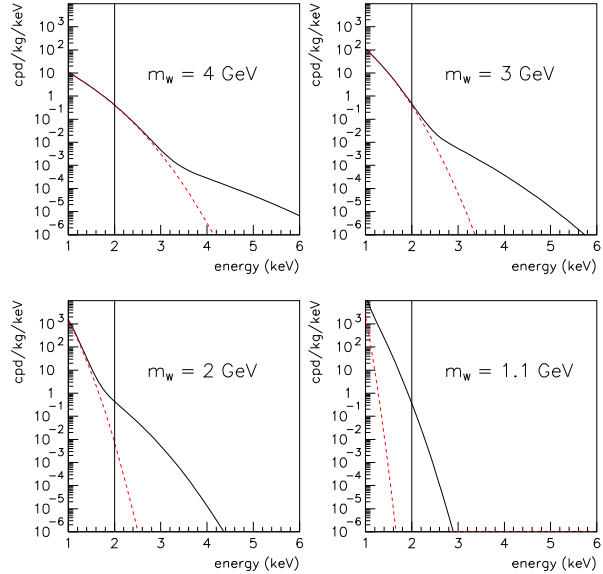}
\vspace{-0.6cm}
\caption{Examples of shapes of expected energy distributions from WIMP-nucleus elastic 
scatterings in the NaI(Tl) detectors of DAMA/NaI with (continuous line) 
and without (dashed line) including the Migdal effect, for the model framework given in the text. 
The effect of the inclusion of this existing physical effect is evident.
The vertical lines indicate the DAMA/NaI energy threshold.}
\label{fg:confro_mg}
\end{figure}

For clarity, Fig. \ref{fg:confro_mg} shows just few examples of shapes of expected energy 
distributions with and without accounting for the Migdal effect.
For this template purpose -- accounting also for the experimental features of the detectors 
\cite{Nim98,Sist,RNC,ijmd} -- we just adopt the following assumptions among all the possibilities:
i) WIMP with dominant Spin Independent coupling and with nuclear cross sections $\propto A^2$;
ii) non-rotating Evans' logarithmic galactic halo model with core radius $R_c = 5$ kpc, 
local velocity $v_0 = 170$ km/s and $\rho_0=0.42$ GeV cm$^{-3}$ (B1 halo model in ref. \cite{RNC,ijmd});
iii) form factors and quenching factors of $^{23}$Na and $^{127}$I
as in case C of ref. \cite{RNC}. The used normalizations assure the same vertical scale in the shown plots. 
It is clear the fraction of events at very low WIMP
masses of electromagnetic nature.
Note that other choices of the model framework do not change the substance of the results.

\section{Some examples}

The proper accounting of the Migdal effect in corollary quests for WIMPs
as DM candidate particles
can be investigated by exploiting  
the expected energy distribution, derived above, to some of the previous analyses 
on the DAMA/NaI annual modulation data
in terms of WIMP-nucleus elastic scattering. For this purpose, the same scaling laws 
and astrophysical, nuclear and
particles physics frameworks of refs. 
\cite{RNC,ijmd} are adopted, while 
-- for simplicity to point out just the impact of the Migdal effect -- 
the SagDEG contribution to the galactic halo,  whose effect we discussed  in ref. \cite{epj06}, 
will not be included here.


The results for each kind of interaction
are presented in terms of allowed volumes/regions,
obtained as superposition of the configurations corresponding
to likelihood function values {\it distant} more than $4\sigma$ from
the null hypothesis (absence of modulation) in each one of the several
(but still a very limited number) of the considered model frameworks.
This allows us to account -- at some extent -- for at least some of the 
existing theoretical and experimental uncertainties  
(see e.g. in  ref. \cite{RNC,ijmd,ijma,epj06} and in the related astrophysics, nuclear and particle
physics literature).

\begin{figure}[!ht]
\centering
\includegraphics[width=150pt] {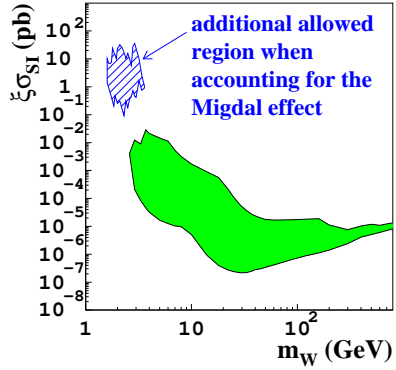}
\vspace{-0.4cm}
\caption{Region allowed in the ($\xi\sigma_{SI},m_W$)
plane in the considered model frameworks 
for pure SI coupling; see text. The hatched region
appears when accounting for the Migdal effect. 
Inclusion of other contributions and/or of other
uncertainties on parameters and models,
such as e.g. the SagDEG contribution \cite{epj06} or more favourable form factors,  
would further extend the region and increases the sets of the best fit values.
For completeness and more see also \cite{RNC,ijmd,ijma,epj06}.}
\label{fg:mwsigsi}
\end{figure}
\begin{figure}[!ht]
\centering
\includegraphics[width=150pt] {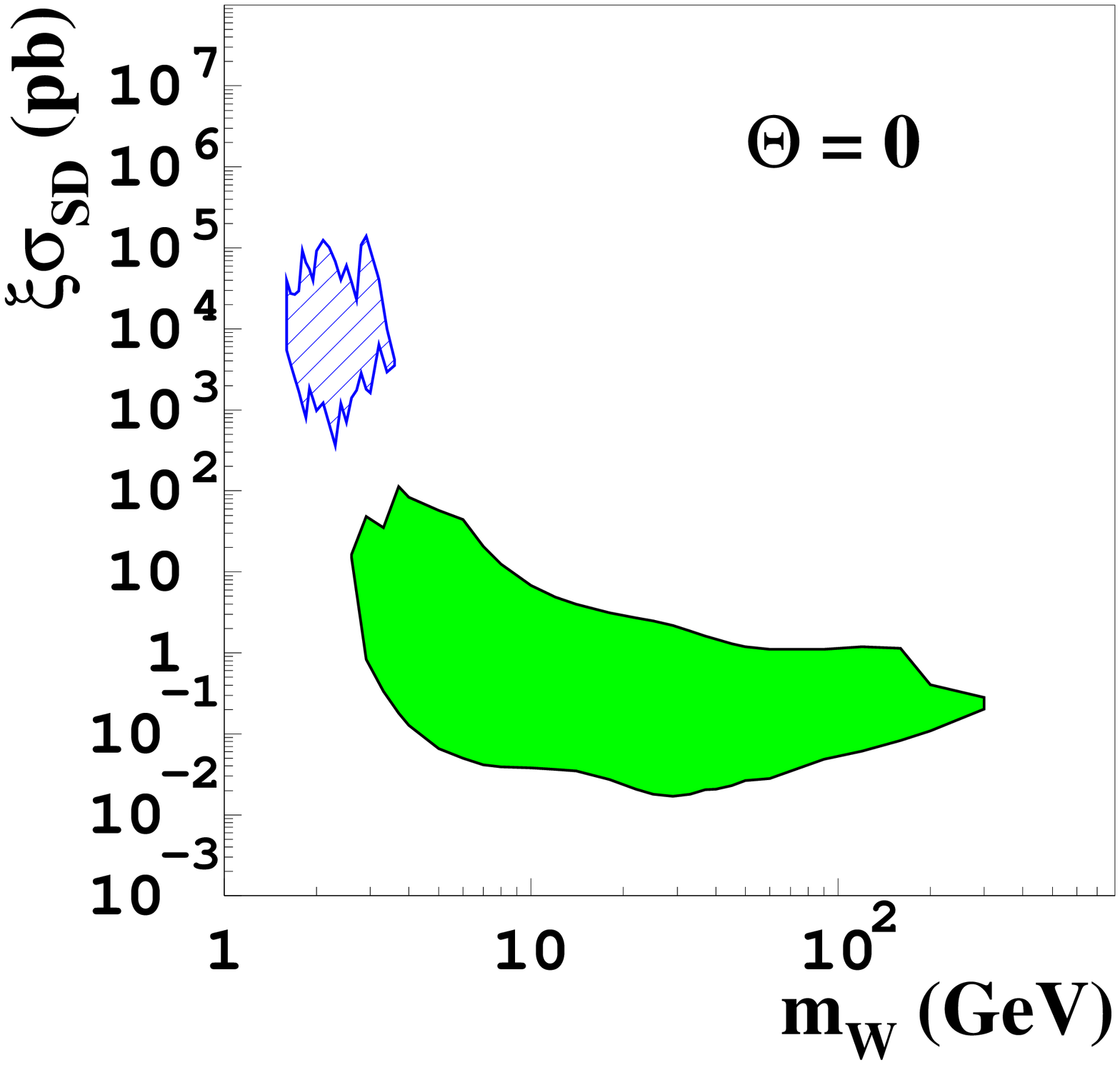}
\includegraphics[width=150pt] {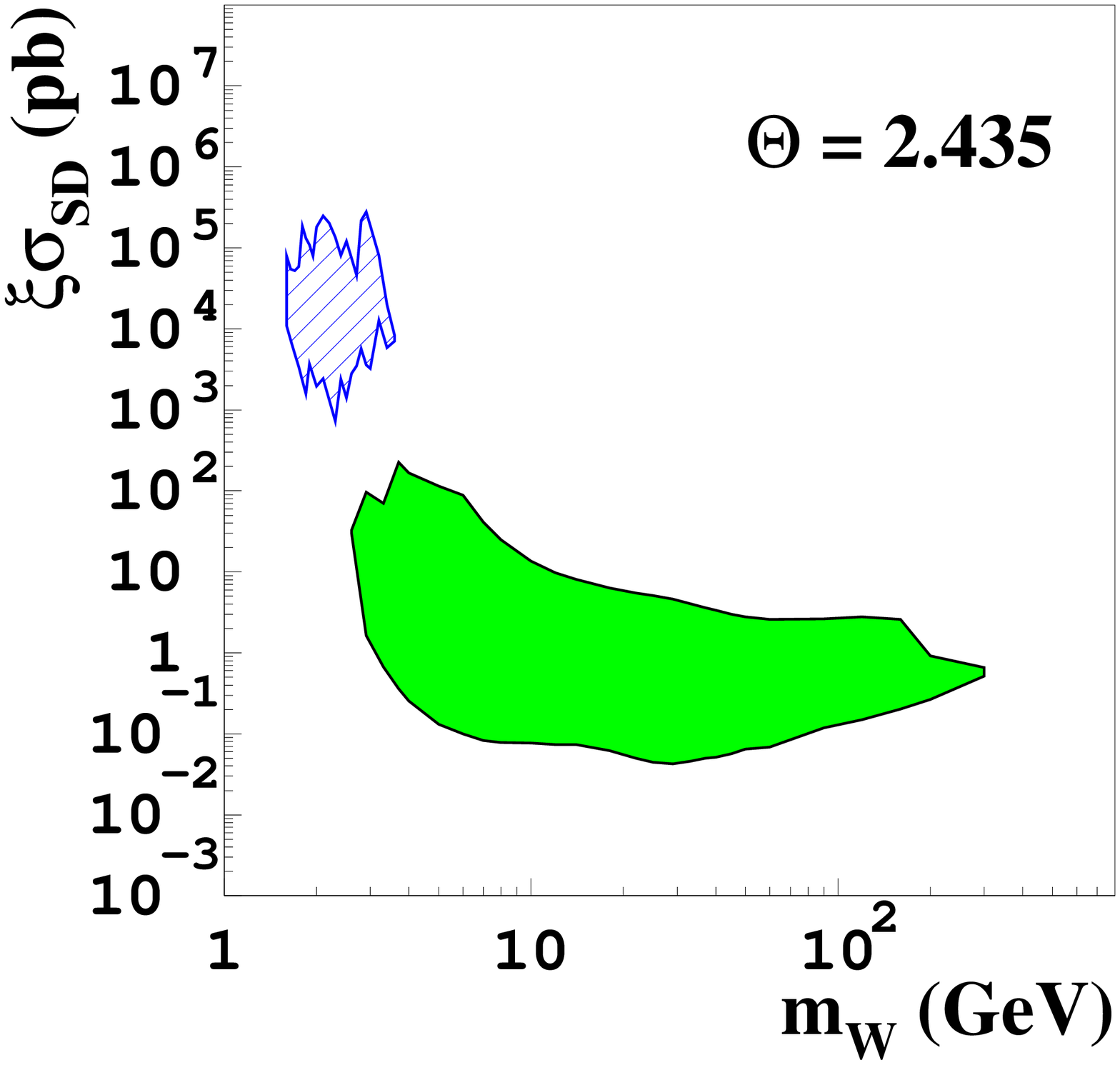}
\vspace{-0.4cm}
\caption{Two slices of the 3-dimensional allowed volume 
($\xi\sigma_{SD},m_W,\theta$)
in the considered model frameworks for pure SD coupling; see text. 
The hatched regions appear when accounting for the Migdal effect. 
Analogous remarks as those in the caption of Fig. \ref{fg:mwsigsi} hold.}
\label{fg:mwsigsd}
\end{figure}

Since the $^{23}$Na and $^{127}$I are fully sensitive both to SI and to SD
interactions, the most general case is defined in a four-dimensional space
($m_W$, $\xi\sigma_{SI}$, $\xi\sigma_{SD}$, $\theta$), where: i) 
$\sigma_{SI}$ is the point-like SI WIMP-nucleon cross section and 
$\sigma_{SD}$ is the point-like SD WIMP-nucleon cross section, according to the definitions and 
scaling laws considered in ref. \cite{RNC}; ii) $tg\theta$ is the ratio between
the effective coupling strengths to neutron and proton for the 
SD couplings ($\theta$ can vary between 0 and $\pi$) \cite{RNC}.
The subcase of purely SI coupled WIMPs is shown in Fig. \ref{fg:mwsigsi},
while in Fig. \ref{fg:mwsigsd} just
two slices of the 3-dimensional allowed volume ($m_W$, $\xi\sigma_{SD}$, $\theta$) for the purely
SD case are given as an example.


It is worth to note that the accounting for the electromagnetic aspects of the interactions provides
in the considered scenarios at the given C.L. additional volumes/regions not topologically connected 
with the remaining allowed parts. 
This depends on the behaviour of the expected energy distributions at low masses (where the Migdal contribution
is appreciable) with respect to that at higher masses, where recoils dominate.

\begin{figure}[!ht]
\centering
\vspace{-0.4cm}
\includegraphics[width=340pt] {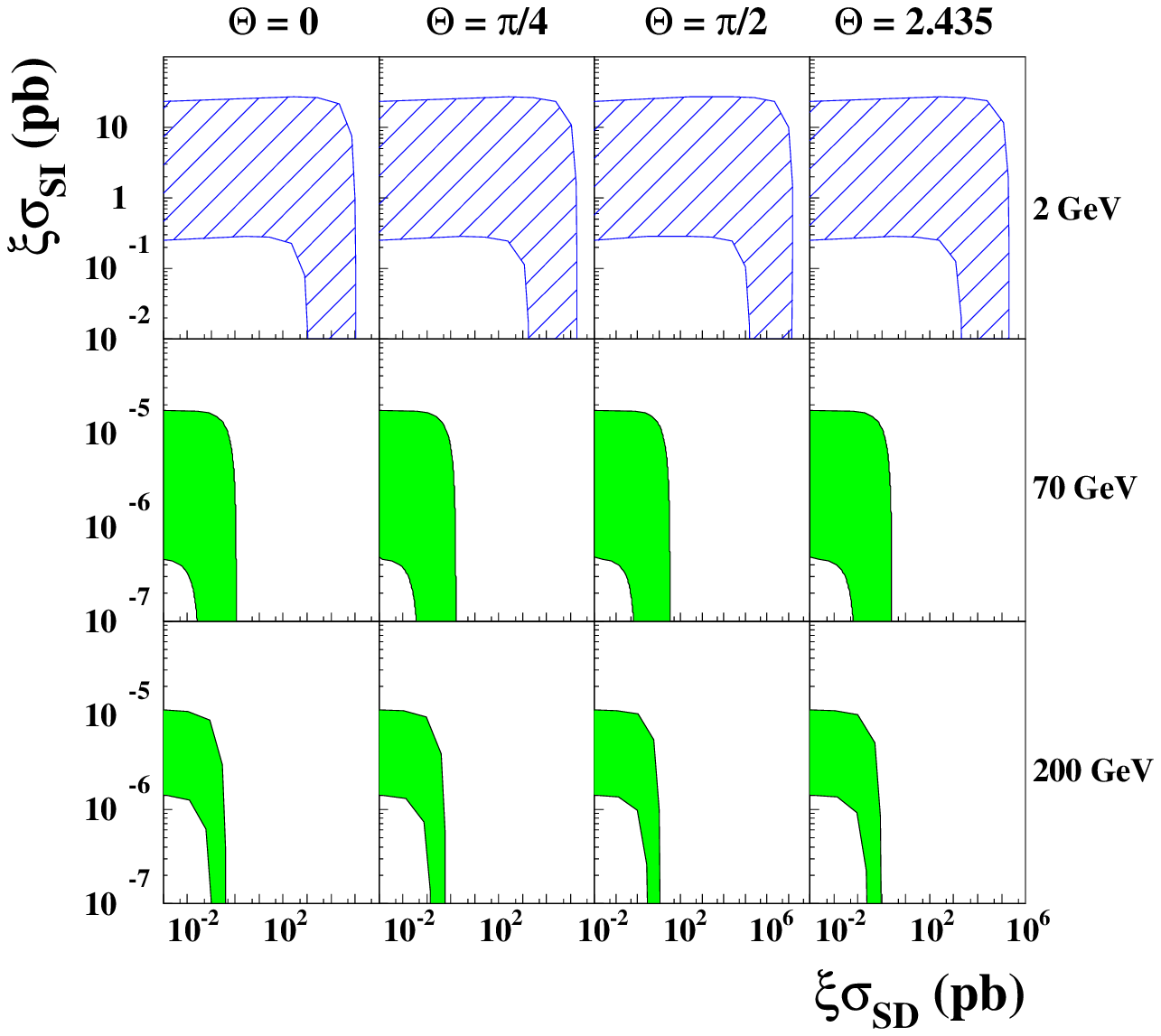}
\vspace{-0.4cm}
\caption{Examples of slices of the 4-dimensional allowed volume 
($\xi\sigma_{SI},\xi\sigma_{SD},m_W,\theta$) in the considered model frameworks; see text. 
The hatched slice appears when accounting for the Migdal effect. 
Analogous remarks as those in the caption of Fig. \ref{fg:mwsigsi} hold.}
\label{fg:sisd}
\end{figure}

Finally, in the general case of mixed SI\&SD coupling 
one gets a 4-dimensional allowed volume ($\xi\sigma_{SI},\xi\sigma_{SD},m_W,\theta$);
new allowed volume at the given C.L. is present in the GeV region.
Fig.\ref{fg:sisd} shows few slices of such a volume as examples.


Note that general comments, extensions and comparisons already discussed in ref. \cite{RNC,ijmd,ijma,epj06}
still hold.

Finally, just for completeness, we remind that
GeV mass DM particles have been proposed in ref. \cite{kitanolow,leptogenesis,farrar}
in order to offer a mechanism able to account for the  
Baryon Asymmetry in the Universe and to naturally 
explain why $\Omega_{DM} \sim 5 \Omega_b$. Moreover,
in ref. \cite{kitanolow} it was shown that a GeV mass DM candidate would potentially solve
the discrepancies between observations and 
$\Lambda CDM$ model on the small scale structure of the Universe.
Finally, among the GeV mass WIMP candidates we remind: 
i) the H dibaryon, already predicted within the Standard Model
of particle Physics \cite{farrar}; ii) the {\em Darkon}, a real scalar field
in an extended Standard Model \cite{he07};
iii) the light photino early proposed  
in models of low-energy supersymmetry \cite{kolb}; iv) 
the very light neutralino in Next-to-MSSM model \cite{NMSSM}; v) 
the scalar GeV mass DM candidates of ref. \cite{fayet}; vi) 
the mirror Deuterium in frameworks where mirror matter interactions  
\cite{zurab} with ordinary matter are dominated by very heavy particles.

\section{Conclusions}

In this paper the ionization and the excitation of bound atomic electrons
induced by the perturbation of the recoiling nucleus after a WIMP elastic
scattering have been discussed. This effect has so far usually been neglected in the field.
The needed theoretical arguments have been developed and the related impact in corollary quests for the 
candidate particle has been shown, as example, for some simplified scenarios.
Obviously, many other arguments can be addressed as well both on DM candidate particles
and on astrophysical, nuclear and particle physics aspects;
for more see \cite{RNC,ijmd,ijma,epj06} and in literature.

\end{document}